\documentstyle[12pt,amsfonts,amssymb,latexsym]{article}

            \newcommand{\be}{\begin{equation}}
            \newcommand{\ee}{\end{equation}}
            \newcommand{\bee}[1]{\begin{equation}\label{#1}}
            \newcommand{\bey}{\begin{eqnarray}}
            \newcommand{\byy}[1]{\begin{eqnarray}\label{#1}}
            \newcommand{\eey}{\end{eqnarray}}
            \newcommand{\R}[1]{(\ref{#1})}
            \newcommand{\C}[1]{\cite{#1}}

            \newcommand{\mvec}[1]{\mbox{\boldmath{$#1$}}}

            \newcommand{\E}{\underline{\bf 1}}

            \newcommand{\ro}{\stackrel{\ \circ}{\varrho}}

            \newcommand{\st}[2]{\stackrel{_#1}{#2}}
            \newcommand{\td}{{^{\bullet}}}
            \newcommand{\dm}{\diamond}
             \newcommand{\D}[1]{\st{\dm}{#1}}

\begin{document}
\date{\empty}
\title{Concepts of Phenomenological Irreversible Quantum 
Thermodynamics I:\\Closed Undecomposed Schottky Systems\\
in Semi-classical Description}
\author{W. Muschik\footnote{Corresponding author:
muschik@physik.tu-berlin.de}
\\
Institut f\"ur Theoretische Physik\\
Technische Universit\"at Berlin\\
Hardenbergstr. 36\\D-10623 BERLIN,  Germany}
\maketitle

\begin{abstract}\noindent
If the von Neumann equation is modified by time dependent statistical weights,
the time rate of entropy, the entropy exchange and production of a Schottky system
are derived whose Hamiltonian does not contain the interaction with the system's
environment. This interaction is semi-classically described by the quantum theoretical
expressions of power- and entropy exchange.
\end{abstract}

\section{Introduction}
Conventional quantum mechanics is a reversible theory because the 
entropy production vanishes for all its processes. There are two 
possibilities to introduce irreversibility into quantum mechanics without 
restricting the full set of
ob\-ser\-vables: One can change Schr\"{o}dinger's equation by introducing
e.g. friction terms, an adventurous procedure which we will not
follow. The se\-cond way is to alter von Neumann's equation by
introducing time dependent weights of the statistical operator, thus
obtaining irreversibility. Other possibilities, introducing a
restricted set of observables thus creating irreversibility by loss of
information \C{MUKA94}, or using statistical concepts as ad-hoc microcanonical
or canonical ensembles, are out of scope of this paper. 
\vspace{.3cm}\newline
An undecomposed closed system interacting with its environment by
heat and power exchange is considered. Undecomposed means, that the
Hamiltonian of the system does not contain an interaction part, neither
for internal interaction nor for that with its environment. As usual for Schottky
systems, the interaction with the environment is phenomenologically described
by heat-, power- and material exchange which is here suppressed considering
closed systems. From a quantum theoretical point of view, this decription is
semi-classical. The task of quantum thermodynamics is to determine entropy
production and exchange quantum-theoretically. This is done by using a modified
von Neumann equation which allows a non-vanishing entropy rate by introducing
time dependent weights of the statistical operator.
\vspace{.3cm}\newline
The paper is organized as follows: After recalling and modifying the von Neumann
equation, the preliminary expressions of entropy exchange and production are
introduced. The entropy exchange contains the contact temperature, a non-equilibrium
analogue of the thermostatic temperature, which is shortly discussed. The consideration
of special processes, such as adiabatic and reversible ones, and of equilibria in isolated
and in closed systems allows to derive a quantum theoretical expression of the
contact temperature.

\section{Schottky Systems\label{UNSY}}

As usual in thermodynamics, we consider a Schottky system
\C{SCHO29,MUASP,MU18}, 
that is a discrete system in interaction with its
environment exchanging heat, power and material. Here, we restrict
ourselves first of all to closed discrete systems for which the
material exchange
is suppressed by a suitable partition between the system and its
environment. The interaction between them is described by the
heat exchange $\st{\td}{Q}$ and by the power exchange $\st{\td}{W}$,
not introducing
a Hamiltonian of interaction. Consequently, we use for the present a 
{\em semi-classical description} of the undecomposed system. The
entropy exchange for this easy case is given by the heat exchange over
the contact temperature $\Theta$, a non- equilibrium temperature which
is defined phenomenologically \C{MU77,MUBR77} and quantum theoretically later on. 
The sum of the entropy exchange and the entropy production is the
entropy time rate of the closed system \C{MU93} whose quantum theoretical
expression is derived by using the Shannon entropy.

\section{The Modified von Neumann Equation}

Starting with a basis of pure quantum states $\{|\Phi^j >\}$ 
which are normalized, complete and orthogonal \C{KATZ67}
\bee{2}
<\Phi^j |\Phi^j >\ =\ 1,\quad \wedge k\neq l:\
<\Phi^k |\Phi^l >= 0,\quad\sum_j |\Phi^j ><\Phi^j |\ =\ \underline{1},
\ee
and  satisfying the  Schr\"{o}dinger equation
\bee{1}
i\hbar\partial_t|\Phi^j >\ =\ {\cal H}|\Phi^j >,\qquad
j=1, 2, 3,... ,
\ee
the self-adjoint non-equilibrium density operator
$\varrho$ is defined by introducing the weights $\{p_j \}$ 
\byy{3}
\varrho\ :=\ \sum_j p_j |\Phi^j ><\Phi^j |,&\quad& 0\leq p_j \leq 1,
\quad \sum_j p_j = 1,\quad\mbox{Tr}\varrho\ =\ 1,
\\ \label{3a}
\varrho |\Phi^k >\! &=&\! p_k  |\Phi^k >.
\eey
From \R{3}$_1$ and \R{1} follows 
\bey\nonumber
\partial_t \varrho = \sum_j \st{\td}{p}_j |\Phi^j ><\Phi^j | +\hspace{4.5cm}
\\ \label{1.1y}
+\sum_j p_j |\frac{1}{i\hbar}{\cal H}\Phi^j ><\Phi^j | + 
\sum_j p_j |\Phi^j ><\frac{1}{i\hbar}{\cal H}\Phi^j |).
\eey
Taking into account that the trace is defined as
\bee{1.1a}
\mbox{Tr}{\cal A}\ =\ \mbox{Tr}\sum_k a_k|\varphi_k><\psi_k|\ =\ \sum_k a_k<\psi_k|\varphi_k>,
\ee
and that the Hamilton operator is self-adjoint, ${\cal H}^+={\cal H}$, we obtain with
\R{2}$_1$ and \R{3}$_3$
\bee{1.1b}
\mbox{Tr}(\partial_t \varrho)\ =\ \sum_{j} \st{\td}{p}_j\ =\ 0,
\vspace{.3cm}\ee
Consequently, the von Neumann equation \R{1.1y}
\bee{5}
\partial_t \varrho\ \equiv\ \st{\td}{\varrho}\ 
=\ -\frac{i}{\hbar}\Big[{\cal H},\varrho\Big]+\ro,\qquad
\ro\ :=\ \sum_j \st{\td}{p}_j|\Phi^j ><\Phi^j |
\ee
results in
\bee{1.4}
\mbox{Tr}(\partial_t \varrho)\ =\ \mbox{Tr}(\ro)\ =\ 0.
\vspace{.3cm}\ee
In contrast to the conventional quantum theory, we introduce the
\vspace{.3cm}\newline
$\blacksquare$\underline{Axiom I:}
\bee{a5}
\vee\ j:\qquad\st{\td}{p}_j\ \neq\ 0\ \longrightarrow\ \ro\ \neq\ 0
\hspace{4cm}\blacksquare\hspace{-2.6cm}
\ee
which generates the {\em propagator} \R{5}$_2$
modifying the von Neumann equation \R{5}$_1$.
\vspace{.3cm}\newline
The modified von Neumann equation shows that the time dependence of the density
operator has two reasons: the quantum mechanical dynamics represented by the 
commutator in \R{5}$_1$ 
and the time dependence of the propagator in
\R{5}$_2$. In conventional quantum theory of isolated systems, the
propagator does not appear,
because the $\{p_j\}$ are presupposed to be time independent, a fact 
which causes reversibility, as we will see below. 
Accepting axiom I means, that irreversibility is generated
by an in time changing composition of the density operator $\varrho$.

\subsection{The first law}

The Hamiltonian  ${\cal H}$ belongs to a non-isolated closed undecomposed system:
no material exchange and missing chemical reactions, but power- and heat-exchange between
system and its  environment. The energy of the considered system is 
\bee{5b}   
E\ :=\ \mbox{Tr}({\cal H}\varrho),
\ee
and the time rate
\bee{5c}
\st{\td}{E}\ =\ \mbox{Tr}(\st{\td}{\cal H}\varrho) + 
\mbox{Tr}({\cal H}\st{\td}{\varrho})
\ee
can be split into power- and heat-exchange according to the 1st 
law of thermodynamics for closed systems
\bee{5d} 
\st{\td}{W}\ :=\ \mbox{Tr}(\st{\td}{\cal H}\varrho),\qquad
\st{\td}{Q}\ :=\ \mbox{Tr}({\cal H}\st{\td}{\varrho}).
\ee
According to \R{5}$_1$, the heat exchange becomes
\bee{5e}
\st{\td}{Q}\ =\ 
\mbox{Tr}\Big({\cal H}(-\frac{i}{\hbar}\Big[{\cal H},\varrho\Big]+\ro)\Big).
\ee
Taking
\bee{5e1}
\mbox{Tr}\Big({\cal H}\Big[{\cal H},\varrho\Big]\Big)\ =\ 
\mbox{Tr}\Big[{\cal H},{\cal H}\varrho\Big]\ =\ 0
\ee
into account, we obtain by \R{5}$_2$ 
\bee{5e2}
\st{\td}{Q}\ =\ \mbox{Tr}\Big({\cal H}\ro\Big)\ =\ 
\mbox{Tr}\Big({\cal H}\sum_j\st{\td}{p}_j|\Phi^j ><\Phi^j |\Big),
\ee
and by introducing suitable work variables $\mathbf{a}$, we obtain
the power exchange \R{5d}$_1$
\bee{5f}
\st{\td}{W}\ =\ \mbox{Tr}\Big(\frac{\partial{\cal H}}{\partial\mathbf{a}}
\varrho\Big)\cdot\st{\td}{\mathbf{a}}\ =:\ \mathbf{A}\cdot\st{\td}{\mathbf{a}} .
\vspace{.3cm}\ee
Power- and energy-rates are exchange quantities between the undecomposed
system and its environment. An interaction Hamiltonian
describing the dependence of the system on its environment
is not introduced. System and environment are not treated as a 
composed (bipartite, compound) system, but their interaction with each other is described 
{\em semi-classically} by
the time dependent composition $\{\st{\td}{p}_j\}$ of the density operator
and by the time rates of the work variables $\st{\td}{\mathbf{a}}$. 
Consequently, the considered system is for the present an undecomposed 
one with an undecomposed Hamiltonian ${\cal H}$. 
Bipartite systems will be treated quantum theoretically in a following paper.

\subsection{Entropy exchange and production}

The non-equilibrium entropy $S$ of the considered system, the Shannon 
entropy, is introduced \C{1KATZ67} as an
\vspace{.3cm}\newline
$\blacksquare$\underline{Axiom II:}
\bee{4}
S(\varrho)\ :=\ -k_B \mbox{Tr} (\varrho\ln\varrho)\hspace{3.5cm}\blacksquare
\hspace{-4cm}
\ee
($k_B\ =\ $ Boltzmann constant).
\vspace{.3cm}\newline
Inserting the modified von Neumann equation \R{5}$_1$ into the Shannon 
non-equilibrium entropy \R{4}, we obtain
\bee{4a}
\st{\td}{S}\ =\ -k_B \mbox{Tr}\Big((-\frac{i}{\hbar}\Big[{\cal H},\varrho\Big]+\ro)\ln\varrho\Big)
-k_B \mbox{Tr}\Big(\varrho\partial_t(\ln\varrho)\Big).
\ee
Inserting
\bee{4b}
\Big[{\cal H},\varrho\Big]\ln\varrho\ =\ {\cal H}\varrho\ln\varrho-\varrho{\cal H}\ln\varrho\ =\
{\cal H}(\ln\varrho)\varrho-\varrho{\cal H}\ln\varrho\ =\
\Big[{\cal H}\ln\varrho,\varrho\Big]
\ee
we obtain
\bee{2.1}
\st{\td}{S}\ =\ -k_B \mbox{Tr} (\ro\ln\varrho)-k_B \mbox{Tr}(\partial_t\varrho)\ =\
 -k_B \mbox{Tr} (\ro\ln\varrho).
\ee
The last term vanishes according to \R{1.4}.
\vspace{.3cm}\newline
Starting with
\bee{2.1a}
 \mbox{Tr}\Big(\ro\ln(Z\varrho)\Big)\ =\ 
\mbox{Tr}\Big(\ro(\ln Z\E+\ln\varrho)\Big)\ =\ \mbox{Tr}(\ro\ln\varrho),
\ee
we obtain by taking \R{1.4} and \R{2.1} into account
\bee{2.3}
\st{\td}{S}\ =\ -k_B \mbox{Tr}\Big(\ro\ln(Z\varrho)\Big),\qquad\wedge Z\in R^1_+.
\ee
Introducing \R{5}$_2$, the entropy rate becomes
\bey\nonumber
\st{\td}{S}(\varrho,\st{\td}{\varrho}) &=&
-k_B \mbox{Tr}\Big[\sum_j\st{\td}{p}_j|\Phi^j><\Phi^j|\ln(Z\varrho)\Big]\ =\
\hspace{4cm}\\ \label{2.3a}
&=&-\sum_j\st{\td}{p}_j<\Phi^j|k_B \ln(Z\varrho)\Phi^j >\ =:\
-\st{\td}{\mathbf p}\cdot{\mathbf f}^I,
\qquad \wedge Z\in R^1_+,
\\ \label{2.3b}
f_k^I &:=& <\Phi^k|k_B \ln(Z\varrho)\Phi^k>.
\eey
The entropy rate $\st{\td}{S}$ does not depend on $Z$ according to \R{2.1a}.
Consequently, $Z$ can be chosen arbitrarily later on. A suitable choice of $Z$ will be
discussed below in connection with equilibrium.
In more detail, we obtain by use of \R{3}$_1$
\bee{7}
\ln(Z\varrho)\ =\ \sum_j\ln(Zp_j)|\Phi^j><\Phi^j|,
\ee
resulting in
\bee{7a}
f_k^I\ =\ \sum_jk_B\ln(Zp_j)\delta^{kj}\ =\ k_B\ln(Zp_k).
\vspace{.3cm}\ee
The entropy rate is thermodynamically decomposed into the entropy exchange 
$\Xi$ and the entropy production $\Sigma$ \C{MU93}
\bee{8}
\st{\td}{S}\ =\ \Xi + \Sigma.
\ee
In this easy case of a closed system,
the entropy exchange is defined by multiplying the heat exchange with the
reciprocal of the contact temperature $\Theta$ which is discussed in sect.\ref{CT}
\bee{a8}
\Xi\ :=\ \frac{\st{\td}{Q}}{\Theta}\ =\
\frac{1}{\Theta}(\st{\td}{E}-{\mathbf A}\cdot\st{\td}{\mathbf{a}})\quad 
\longrightarrow\quad \st{\td}{S}\ =\ \Sigma
+ \frac{1}{\Theta}(\st{\td}{E}-\mathbf{A}\cdot\st{\td}{\mathbf{a}})
\ee
according to the first law \R{5c} and \R{8}. 
Using \R{5e2}, we obtain for the entropy exchange in the case of a non-isolated
closed system
\bey\nonumber
\Xi\ &=&
\mbox{Tr}\Big(\frac{{\cal H}}{\Theta}\sum_j\st{\td}{p}_j|\Phi^j>
<\Phi^j |\Big)\ =
\\ \label{9} 
&=& \sum_j\st{\td}{p}_j<\Phi^j |\frac{{\cal H}}{\Theta}\Phi^j >\ =:\ 
\st{\td}{\mathbf p}\cdot{\mathbf f}^{II},\quad f^{II}_k\ :=\ 
<\Phi^k |\frac{{\cal H}}{\Theta}\Phi^k >.\hspace{.5cm}
\eey
Because the entropy rate \R{2.3} and the entropy exchange \R{9} do not
depend on the parameter $Z$, also the entropy production is independent of it.
As we demonstrate below, $Z$ is chosen differently for isolated and for closed
non-isolated sytems.
\vspace{.3cm}\newline
The exchange quantites $\st{\td}{Q}$ and $\Xi$ depend on the state of the system 
and on that of its environment \C{MUBE04}. Consequently, $\st{\td}{\mathbf p}$
contains a part 
which is determined by the environment, as we will see below in more detail.
The thermodynamic quantities, entropy rate \R{2.3a}, entropy exchange \R{9} and
entropy production require a modified von Neumann equation according
to \R{a5}, demonstrating as expected: conventional quantum theory with 
$\st{\td}{\mathbf p}\equiv\mvec{0}$ is a non-thermal theory.\nolinebreak
\vspace{.3cm}\newline
According to \R{1}, \R{3}$_1$ and \R{5f}, the Hamilton operator $\cal H$ and the
density operator $\varrho$ do not depend on $\st{\td}{\mathbf p}$ and
$\st{\td}{\mathbf{a}}$. Consequently, also the quantum theoretical expressions
$\mathbf f^I$ and $\mathbf f^{II}$ in \R{2.3a} and \R{9} are idependent of
$\st{\td}{\mathbf p}$ and $\st{\td}{\mathbf{a}}$, and the interaction between
system and environment is semi-classically described by $\rho$, that means by
$\st{\td}{\mathbf p}$ and $\st{\td}{\mathbf{a}}$.
\vspace{.3cm}\newline
According to \R{8}, \R{2.3a} and \R{9}, the entropy production results in
\bee{10}
\Sigma\ =\ \st{\td}{S}-\Xi\ =\ -\st{\td}{\mathbf p}\cdot({\mathbf f^I+\mathbf f^{II}})\
=:\ -\st{\td}{\mathbf p}\cdot{\mathbf f}\ \stackrel{_*}{\geq}\ 0,
\ee
The entropy production is an internal quantity of the system, that means, it does
not depend on the exchange quantities between system and environment, here
in semi-classical description:
\vspace{.3cm}\newline
$\blacksquare$\underline{Axiom III:}\newline
Power- and heat exchange, $\st{\td}{W}(t)$ and $\st{\td}{Q}(t)$, are independent of
each other and do not influence the entropy production $\Sigma(t)$ locally in time.  
\hfill$\blacksquare$

\section{Contact Temperature\label{CT}}

The non-equilibrium contact temperature $\Theta$  in \R{a8} is defined by
the inequality \C{MU77,MUBR77,MUBE07,MU09,MU18a} 
\bee{8a}
\st{\td}{Q}\Big(\frac{1}{\Theta}-\frac{1}{T^\Box}\Big)\ \geq\ 0
\ee
as follows: the system is contacted with an equilibrium environment
of the thermostatic temperature $T^\Box$ generating the net heat exchange
$\st{\td}{Q}$. For defining the contact temperature, we choose a special
equilibrium environment  so that the net heat exchange between system and
environment vanishes. That occurs, if the environment has the thermostatic
temperature $T^\Box_\odot$. Now using the
\vspace{.3cm}\newline
$\blacksquare$\underline{Proposition} \C{MU85}
\bey\nonumber
{\bf X}\cdot f({\bf X})\ \geq\ {\bf 0}\ (\mbox{for all}\ {\bf X}\wedge f \ \mbox{continuous at}\ 
{\bf X} = {\bf 0})\ \Longrightarrow\ 
\\ \nonumber
\Longrightarrow\ f({\bf 0}) = {\bf 0}\ \longleftrightarrow\  f({\bf X})\ =\ 
 {\bf M}({\bf X})\cdot{\bf X},
\\ \label{9a}
 {\bf M}({\bf X})\ \mbox{positive semi-definite},
\hspace{.8cm}\blacksquare\hspace{-1cm}
\eey
a comparison with \R{8a} shows that 
\bee{9b}
\st{\td}{Q}\ =\ 0\ \longleftrightarrow\ \Theta=T^\Box_\odot.
\ee
In more detail:
\begin{center}
\parbox[t]{11cm}{
{\sf Definition:} The system's contact temperature is that thermostatic temperature
of the system's
equilibrium environment for which the net heat exchange between the system and
this environment
through an inert partition vanishes by change of sign.} 
\end{center}
We need the non-equilibrium contact temperature $\Theta$ 
because it is a quantity belonging to
the system, a state function. A quantum theoretical definition of the contact
temperature is given below. 
\vspace{.3cm}\newline
As easily to demonstrate, contact temperature $\Theta$ and the internal energy $U$ are
independent of each other. For this purpose, a rigid inert partition 
($\st{\td}{\mvec{a}} \equiv\mvec{0}$) is chosen which is impervious to matter
$(\st{\td}{\mvec{n}}\!{^e}\equiv 0)$ and a time-dependent environment temperature
$T^\Box(t)$ which is always set equal to the value of the momentary contact temperature
$\Theta(t)$ of the closed system: 
\bee{33}
T^\Box (t)\st{*}{=}\Theta (t)\ \longrightarrow\ \st{\td}{Q}\!{^\Box} =
-\st{\td}{Q}\ =\ 0\ \longrightarrow\ \st{\td}{U}\ =\ 0
\ee
according to \R{9b} and an inert partition between system and environment.
Because $\Theta$ is time-dependent and $U$ is constant, totally
different from thermostatics, both quantities are independent of each other.

\section{Special Processes\label{SP}}

In this section, we consider special thermodynamical
processes and their quantum theoretical interpretations. The well-known 
phenomenological concepts of adiabatic, 
irreversible and reversible processes in isolated and in closed
systems and the concept of equilibrium are interpreted 
quantum-theoretically without using methods of statistical thermodynamics: 
we are looking for a {\em phenomeno\-lo\-gical irreversible quantum 
thermodynamics}.

\subsection{Isolated systems\label{IS}}

Isolated systems are phenomenological defined by a partition between 
system and its environment which is
impervious to heat, power and material. That means, heat-, power- and 
material-exchange vanish identically for arbitrary
states of system and environment. Because the particle number 
operator was not taken into account in the energy rate \R{5c}, the systems
considered here are of constant particle number. 
Consequently, the particle
number is not a variable of the system, and therefore the material exchange 
between system and environment
vanishes identically and chemical reactions are absent.
Such systems are denoted as {\em closed systems
without chemical reactions}.
\vspace{.3cm}\newline
{\em Isolated} systems are closed systems with identically vanishing 
heat-, power- and entropy-exchange. 
According to \R{5f} and \R{9}, 
we have the following conditions of isolation
\bee{41}  
{\st{\td}{\mathbf{a}}}{^{iso}}\ \equiv\ {\mathbf{0}}\ \longrightarrow\
\st{\td}{W}_{iso}\ \equiv\ 0,\qquad \st{\td}{Q}_{iso}\ \equiv\ 0\ 
\longrightarrow\ \Xi_{iso}\ \equiv\ 0.
\ee
These conditions are achieved by introducing an isolating partition between
system and environment which does not influence the state of the system.
Thus we accept the following
\vspace{.3cm}\newline
$\blacksquare$\underline{Axiom IV:}\newline
The introduction of an isolating partition between system and environment
does not change the Hamiltonian and the density operator of an undecomposed 
system in semi-classical description. This isolation influences the time rate of the
density operator by changing the time rates of its weights
\bee{x41}
\st{\td}{\mathbf p}\quad \longrightarrow\quad  \st{\td}{\mathbf p}\!{^{iso}}.
\hspace{4.5cm}\blacksquare\hspace{-4cm}
\vspace{.3cm}\ee
Evident is that the exchange quantites $\st{\td}{W}$ and $\st{\td}{Q}$ depend
also on the state of the system's environment. According to \R{5f}, the power
exchange is controlled by the rates of the work variables $\st{\td}{\mathbf{a}}$,
whereas the heat exchange \R{5e2}$_2$ is represented by the rates of the weights
$\st{\td}{\mathbf p}$ because the Hamiltonian does not depend on
$\st{\td}{\mathbf{a}}$ and does not contain any interaction with the environment
in semi-classical description. Consequently, when isolating the system from its 
environment, these rates change according to axiom IV \R{x41}.
\vspace{.3cm}\newline
Introducing
\bee{a41}
\st{\td}{\mathbf{p}}\!{^{ex}}\ :=\ \st{\td}{\mathbf{p}}\ -\ \st{\td}{\mathbf{p}}\!{^{iso}},
\quad\longrightarrow\quad\st{\td}{\mathbf{p}}\!{^{ex,iso}}\ =\ \mathbf{0},
\ee
we obtain for the time rates of the weights in
\byy{a41a}
\mbox{non-isolated closed systems}&\qquad&
\st{\td}{\mathbf{p}}\ =\ \st{\td}{\mathbf{p}}\!{^{iso}}
+\st{\td}{\mathbf{p}}\!{^{ex}},
\\ \label{a41b}
\mbox{isolated systems}&\qquad&\st{\td}{\mathbf{p}}\!{^{iso}}.
\eey
The quantities ${\mathbf f}^I$ and ${\mathbf f}^{II}$, \R{2.3b} and \R{9}$_4$,
are independent of an isolation of the system.
\vspace{.3cm}\newline
According to \R{9}$_3$ and \R{41}$_4$, we obtain for an isolated system
\bee{40g}
\Xi_{iso}\ \equiv\ 0\ =\ \st{\td}{\mathbf p}\!{^{iso}}\cdot{\mathbf f}^{II},
\qquad \sum_j\st{\td}{p}_j\!\!^{iso}\ =\ 0.
\ee
From \R{5}$_2$ we obtain the split into an {\em exchange propagator} 
$\ro_{ex}$ which can be quantum theoretically identified later
on, and a dissipative thermal part  $\ro_{iso}$, called the 
{\em irreversibility propagator}, 
\byy{41d}
\ro\ = \ \ro_{ex} + \ro_{iso},\hspace{4.6cm}
\\ \label{41e}
\ro_{ex}\ =\ \sum_j \Big\{\st{\td}{p}_j\!{^{ex}}|\Phi^j><\Phi^j|\Big\},\qquad
\ro_{iso}\ =\ \sum_j \Big\{\st{\td}{p}_j\!{^{iso}}|\Phi^j><\Phi^j|\Big\}.
\eey
Because of \R{1.1b}$_2$ and \R{40g}$_2$, we obtain
\bee{a41e}
\sum_j\st{\td}{p}_j\!{^{ex}}\ =\ 0,
\quad\longrightarrow\quad{\mathbf{e}\cdot\st{\td}{\mathbf{p}}}{^{ex}}\ =\ 0,\quad
{\mathbf{e}\cdot\st{\td}{\mathbf{p}}}{^{iso}}\ =\ 0,\quad e_j=1,
\ee
and the traces of both parts of the propagator vanish
\bee{41f}
\mbox{Tr}\ro_{ex}\ =\ 0,\qquad\mbox{Tr}\ro_{iso}\ =\ 0.
\vspace{.3cm}\ee
According to \R{9}$_3$ and \R{40g}$_1$, the entropy exchange becomes
\bee{M1}
\Xi\ =\ \st{\td}{\mathbf p}\!{^{ex}}\cdot{\mathbf f}^{II},
\ee
and the entropy production \R{10} yields
\bee{M2}
\Sigma\ =\ -\st{\td}{\mathbf p}\cdot{\mathbf f}^{I}
-\st{\td}{\mathbf p}\!{^{ex}}\cdot{\mathbf f}^{II}\ =\ 
-\st{\td}{\mathbf p}\!{^{iso}}\cdot{\mathbf f}^{I}
-\st{\td}{\mathbf p}\!{^{ex}}\cdot{\mathbf f}^{I}
-\st{\td}{\mathbf p}\!{^{ex}}\cdot{\mathbf f}^{II}.
\ee
According to \R{2.3a}$_3$, the entropy rate in an isolated system is
\bee{M3}
\st{\td}{S}_{iso}\ =\ -\st{\td}{\mathbf p}\!{^{iso}}\cdot{\mathbf f}^{I},
\ee
resulting \R{M2} in
\bee{M4}
\Sigma\ =\ \st{\td}{S}_{iso}- \st{\td}{\mathbf p}\!{^{ex}}\cdot{\mathbf f}.
\ee
Taking axiom III into consideration,
entropy production and entropy rate are connected by the following
\vspace{.3cm}\newline
$\blacksquare$\underline{Axiom V:}\newline
The entropy production of a non-isolated closed system is defined as the entropy 
rate of the same system in isolation
\bee{bb41f} 
\Sigma\ :=\ \st{\td}{S}_{iso}\ \longrightarrow\ 
\st{\td}{\mathbf p}\!{^{ex}}\cdot{\mathbf f}\ =\ 0\ \longrightarrow\ 
\st{\td}{\mathbf p}\!{^{ex}}\cdot{\mathbf f}^{II}\ =\ 
-\st{\td}{\mathbf p}\!{^{ex}}\cdot{\mathbf f}^I
\ee
according to \R{M4}.
\mbox{}\hfill$\blacksquare$\vspace{.3cm}\newline
According to \R{10}$_3$, \R{M3} and \R{bb41f}$_1$, we obtain
\bee{bb41g}
\Sigma\ =\ -\st{\td}{\mathbf p}\cdot{\mathbf f}\ =\ 
-\st{\td}{\mathbf p}\!{^{iso}}\cdot{\mathbf f}^{I}\ \geq\ 0.
\ee
Taking \R{2.3a}$_3$, \R{bb41g}$_2$, \R{bb41f}$_3$  and \R{M1} into account,
the entropy rate becomes
\bee{M5}
\st{\td}{S}\ =\ \st{\td}{\mathbf p}\!{^{ex}}\cdot{\mathbf f}^{II}\ 
-\st{\td}{\mathbf p}\!{^{iso}}\cdot{\mathbf f}^{I},
\ee
and we realize the different meanings
of $\ro_{ex}$ and $\ro_{iso}$: the irreversibility propagator $\ro_{iso}$ belongs
to the entropy production $\Sigma$ according to \R{M3} and \R{bb41f}$_1$,
whereas the exchange propagator $\ro_{ex}$ belongs to the entropy exchange
$\Xi$ according to \R{M1}.

\subsection{Adiabatic processes}

If an isolated system is opened for power exchange, it performs an 
{\em adiabatic process} which is characterized by
\bee{42}
\st{\td}{Q}_{ad}\ =\ 0\ \longrightarrow\ \Xi_{ad}\ =\ 0\ 
\wedge\ \st{\td}{E}_{ad}\ =\ {\mathbf{A}}\cdot{\mathbf\st{\td}{a}}\ \neq\ 0
\ee
according to \R{a8}$_1$. As in isolated systems \R{41}, the heat exchange
 vanishes in adiabatic isolated systems, whereas the rates of the
work variables are different from zero. Because the entropy production does not depend
on the rates of the work variables ${\mathbf\st{\td}{a}}$ according to axiom III,
the entropy rate is the same in isolated and adiabatic systems according to
\R{a8}$_2$, a fact which is also true for the vanishing entropy exchanges.
Consequently, the thermal statements are identical for isolated and adiabatic systems.

\subsection{Reversible processes and quantum theory}

According to \R{bb41g}, {\em reversible processes} in are defined by
\bee{51x}
\Sigma_{rev}\ =\ -\st{\td}{\mathbf p}_{rev}\cdot{\mathbf f}\ =\ 
-\st{\td}{\mathbf p}\!{^{iso}_{rev}}\cdot{\mathbf f}^{I}\ =\ 0.
\ee
The "time" in the rates of \R{51x}  
is the path parameter along the reversible process on the equilibrium sub-space.
According to \R{51x}, we have to distinguish between reversible processes in
\byy{51y}
\mbox{non-isolated systems:}&\qquad&\st{\td}{\mathbf p}_{rev}\cdot{\mathbf f}\ =\ 0,
\quad{\mathbf f} \neq {\bf 0},\ \st{\td}{\mathbf p}_{rev} \neq {\bf 0},
\\ \label{51z}
\mbox{isolated systems:}&\qquad&\st{\td}{\mathbf p}\!{^{iso}_{rev}}\cdot{\mathbf f}^{I}\ =\ 0,\quad{\mathbf f}^{I} \neq {\bf 0},\ \st{\td}{\mathbf p}\!{^{iso}_{rev}}
\neq {\bf 0}.
\eey
In {\em coventional quantum mechanics} of undecomposed closed systems, the
$\st{\td}{\mathbf{p}}$ and therefore $\ro$ are zero, and according to \R{a41}
\bee{42a} 
\st{\td}{\mathbf{p}}\!{^{ex}_{qu}}\ =\ -\st{\td}{\mathbf{p}}\!{^{iso}_{qu}}
\ee
is valid. According to \R{M1}, \R{bb41f}$_3$, \R{42a} and \R{bb41g}$_2$,
we obtain for an (adiabatic) process in conventional quantum mechanics
\bee{42b}
0\ =\ \Xi_{ad}\ =\ -\st{\td}{\mathbf{p}}\!{^{ex}_{ad}}\cdot{\mathbf f}^I\ =\
\st{\td}{\mathbf{p}}\!{^{iso}_{qu}}\cdot{\mathbf f}^I\ =\ -\Sigma_{qu}\ 
\longrightarrow\ \st{\td}{S}_{qu}\ =\ 0.
\ee
Consequently, entropy exchange, -production and -rate
vanish in conventional quantum theory. That is the reason, why it
is regarded as a reversible theory of adiabatic processes.

\section{Equilibria\label{EQUI}}

Reversible processes are defined as trajectories on the equilibrium sub-space, that
means, a reversible process consists of equilibrium states which are defined according
to \R{51y} and \R{51z} by the following {\em equilibrium conditions}
\byy{45}
{\mathbf{\st{\td}{a}}}{^{eq}}\ \doteq\ {\mathbf{0}} \ &\wedge&\
\st{\td}{\varrho}_{eq}\ \doteq\ 0,
\\ \label{45A}
\st{\td}{\mathbf{p}}{_{eq}}
\doteq\ {\mathbf{0}}\ \ \wedge\ \st{\td}{\mathbf{p}}{^{iso}_{eq}}
\doteq\ \mathbf{0}\ &\wedge&\ {\mathbf f}^{eq}\ \doteq\ {\bf 0}
\ \wedge\ {\mathbf f}^{Ieq}\ \doteq\ {\bf 0}.
\eey
According to \R{5f}, \R{5} and \R{5e2}, \R{2.3}, \R{9} and \R{10}, 
we obtain from \R{45} and \R{45A}$_{1,2}$
\byy{a45}
\st{\td}{W}_{eq}\ =\ 0,\quad\ro_{eq}\ =\ 0,\quad [{\cal H},\varrho_{eq}]\ =\ 0, 
\\ \label{a451}\st{\td}{Q}_{eq}\ =\ 0,\quad
 \st{\td}{S}_{eq}\ =\ 0,\quad \Xi_{eq}\ =\ 0,\quad\Sigma_{eq}\ =\ 0.
\vspace{.3cm}\eey
According to \R{45A}$_{3,4}$, we have to distinguish between to different kinds of
equilibria: equilibrium in non-isolated systems according to \R{45A}$_{3}$,
\R{2.3b} and \R{9}$_3$
\bee{c45} 
f_j^{eq}\  =\ f_j^{Ieq} + f_j^{IIeq}\ =\
<\Phi^j |\Big(k_B \ln(Z\varrho_{eq}) 
+ \frac{{\cal H}}{\Theta_{eq}}\Big)\Phi^j >\ =\ 0,\quad\wedge j, 
\ee
and equilibrium in isolated systems according \R{2.3b} and \R{7a}
\bee{d45}
f_k^{Ieq}\ =\ <\Phi^k_{eq} |k_B \ln(Z\varrho_{eq})\ \Phi^k_{eq} >\ 
=\ 0\ =\ k_B\ln(Zp_k^{eq}). \quad\wedge k.
\ee

\subsection{Isolated systems}

From \R{d45}$_3$ follows
\bee{e45}
Zp_k^{eq}\ =\ 1\ \longrightarrow\ p_k^{eq}\ =\ \frac{1}{Z}\ 
\longrightarrow\ \varrho_{eq}\ =\ \frac{1}{Z}\sum_k|\Phi^k><\Phi^k|\ =\ 
\frac{1}{Z}{\underline 1}.
\ee
Tracing the density operator according to \R{3}$_4$ results in
\bee{5.1a}
1\ =\ \sum_j <\Phi^j|\varrho_{eq}\Phi^j>\ 
=\ \frac{1}{Z}\sum_j<\Phi^j|\Phi^j>\ =\ \frac{1}{Z}\sum_j 1_j.
\ee
The last sum must be restricted because of convergence: $1\leq j\leq N$. 
Consequently according to \R{5.1a}$_3$
$Z=N$ is valid, and the density operator \R{e45}$_3$ of an 
isolated system has, as expected, the micro-canonical form
\bee{5.1bx}
\varrho_{mic}\ =\ \frac{1}{N}\sum_{j=1}^N |\Phi^j><{\Phi}^j|,\qquad
N<\infty. 
\ee

\subsection{Non-isolated closed systems}

According to \R{a45}$_3$, the Hamilton operator commutes with the density ope\-rator
in equilibrium. Consequently, a common system of eigenfunctions exists for both
operators, and we presuppose that this system is given by \R{3a}. Consequently, from
\R{c45}$_3$ follows
\bey\nonumber
 k_B\ln(Zp_k^{eq}) + \frac{E_k}{\Theta_{eq}}\ =\ 0\ \longrightarrow\hspace{6.5cm}
\\ \label{I3}
\longrightarrow\ k_B
\sum_k\ln(Zp_k^{eq})|\Phi^k><{\Phi}^k| 
+\sum_k |\Phi^k>\frac{E_k}{\Theta_{eq}}<{\Phi}^k|\ =\ 0,
\eey 
resulting according to \R{7} in
\bee{I4}
k_B\ln(Z\varrho_{eq}) + \frac{\cal H}{\Theta_{eq}}\ =\ \underline{0}.
\ee
From \R{I4} follows the canonical density operator for a closed 
non-isolated system in equilibrium
\bee{a48}
\varrho_{can}\ =\ \frac{1}{Z} 
\exp\Big[-\frac{{\cal H}}{k_B \Theta}_{eq}\Big],\quad
Z\ =\ \mbox{Tr}\exp\Big[-\frac{{\cal H}}{k_B \Theta}_{eq}\Big].
\vspace{.3cm}\ee
If we presuppose that the equilibrium environment which in contact with the 
system is a heat reservoir of the thermostatic temperature $T^\Box_\odot$,
the equilibrium contact temperature $\Theta_{eq}$ of the system 
is replaced by $T^\Box_\odot$
\bee{48a}
\Theta_{eq}\ =\ T^\Box_\odot,
\ee
representing an additional equilibrium condition.
\vspace{.3cm}\newline
The micro-canonical and the canonical equilibrium density operators, \R{5.1bx} and
\R{a48}, are derived by a pure phenomenological argumentation: starting with the
entropy production in isolated and non-isolated systems \R{bb41g}, reversible
processes \R{51x} are defined by vanishing entropy production, a condition which is
also satisfied for equilibria. But for equilibria, the vanishing entropy production
follows from the equilibrium conditions \R{45} and \R{45A}, thus distinguishing
reversible processes from equilibria. The equilibrium density operators $\varrho_{mic}$
\R{5.1bx} and $\varrho_{can}$ \R{a48} follow from the phenomenological equilibrium
conditions without using the 2nd law and a quantum theoretical background. A
comparison of \R{51y} and \R{51z} with \R{45A} shows that the entropy production
in equilibrium vanishes with a "higher grade" than that for reversible processes.

\section{Constitutive Equations of the Propagators}

In conventional quantum theory, the constitutive properties of the considered system
are described by the Hamiltonian which is not equipped with thermal properties in
semi-classical description. Here, the constitutive properties are introduced by the
propagators \R{41e}. 
\vspace{.3cm}\newline
As in classical non-equilibrium thermodynamics, the entropy production 
in quantum thermodynamics has also the typical form of a product of 
``fluxes'' $\st{\td}{\mathbf{p}}$ and ``forces'' $\mathbf{f}$ according to \R{bb41g}.
The term $\st{\td}{\mathbf p}{^{iso}}\cdot{\mathbf f}^I$ belongs to an isolated
system, whereas the term $\st{\td}{\mathbf p}\cdot{\mathbf f}$ to a non-isolated
closed system. According to the equilibrium conditions \R{45A}, we obtain the
non-linear constitutive equations
\byy{C1}
\st{\td}{\mathbf p}{^{iso}} &=&{\bf B}({\mathbf f}^I)\cdot{\mathbf f}^I,
\qquad\mbox{${\bf B}^\top={\bf B}$, negative definite, singular},
\\ \label{C2}
\st{\td}{\mathbf p}{^{ex}} &=&{\bf A}({\mathbf f},\Xi)\cdot{\mathbf f},\qquad
{\bf A}^\top = -{\bf A},\qquad{\bf A}({\mathbf f},0)\ =\ {\bf 0}.
\eey
These constitutive equations satisfy the relevant relations of chap.\ref{EQUI} and
\ref{SP}, as we will demonstrate below.
\vspace{.3cm}\newline
According to \R{M1} and \R{C2}, the entropy exchange becomes
\bee{C2a}
\Xi\ =\ {\mathbf f}^{II}\cdot{\bf A}({\mathbf f},\Xi)\cdot{\mathbf f},
\ee
and  the following relations are valid according to \R{bb41g}$_2$, \R{40g}$_1$,
and \R{bb41f}$_2$,
\bee{C3}
{\mathbf f}^I\cdot{\bf B}({\mathbf f}^I)\cdot{\mathbf f}^I\ \leq\ 0,\quad
{\mathbf f}^{II}\cdot{\bf B}({\mathbf f}^I)\cdot{\mathbf f}^I\ =\ 0,\quad
{\mathbf f}\cdot{\bf A}({\mathbf f},\Xi)\cdot{\mathbf f}\ =\ 0.
\ee
According to \R{10}$_{3,4}$, the negative entropy production is by use of \R{C1},
\R{C2}, \R{a41}$_1$, \R{C3} and \R{10}$_3$
\bey\nonumber
-\Sigma\ =\ 
\st{\td}{\mathbf p}\cdot{\mathbf f} &=& {\mathbf f}\cdot\Big({\bf A}\cdot{\mathbf f}+
{\bf B}\cdot{\mathbf f}^I\Big)\ =\ {\mathbf f}\cdot{\bf B}\cdot{\mathbf f}^I\ =
\\ \label{C5}
&=&\Big({\mathbf f}^I + {\mathbf f}^{II}\Big)\cdot{\bf B}\cdot{\mathbf f}^I\ =\ 
{\mathbf f}^I\cdot{\bf B}\cdot{\mathbf f}^I\ \leq\ 0.
\vspace{.3cm}\eey
We now consider the equilibria in isolated systems
\bee{C6}
{\mathbf f}^I ={\bf 0}\ \longrightarrow\ 
\st{\td}{\mathbf p}{^{iso}_{eq}}={\bf 0}\ \wedge\ {\mathbf f}={\mathbf f}^{II}\  
\longrightarrow\ \Xi=0\ \longrightarrow\ \st{\td}{\mathbf p}{^{ex}_{eq}}={\bf 0}, 
\ee
and in non-isolated closed systems
\bee{C7}
{\mathbf f} = {\bf 0}\ \longrightarrow\ \st{\td}{\mathbf p}{^{ex}_{eq}}={\bf 0}\ 
\wedge\ \Xi=0\ \wedge\ {\mathbf f}^I =-{\mathbf f}^{II}\ \longrightarrow\ 
-{\bf B}\cdot{\mathbf f}^{II}=\ \st{\td}{\mathbf p}{^{iso}_{eq}}\doteq {\bf 0}.
\ee
The setting
\bee{C8}
{\bf B}\cdot{\mathbf f}^{II}\ \doteq\ {\bf 0} 
\ee
is compatible with \R{C3}$_2$. Thus, the constitutive equations \R{C1} and \R{C2}
are in accordance with the constraints \R{C3}, with the entropy production \R{C5} and
with the equilibria \R{C6} and \R{C7} in isolated and non-isolated closed systems.
\vspace{.3cm}\newline
Taking \R{C1} and \R{C2} into account, the modified von Neumann equation \R{5}$_1$
writes with \R{41d} and \R{41e}
\bee{C9}
\st{\td}{\varrho}\ 
=\ -\frac{i}{\hbar}\Big[{\cal H},\varrho\Big]+
\sum_{jk}\Big(B_{jk}f^I_k + A_{jk}f_k\Big)|\Phi^j><\Phi^j|.
\ee
The last term becomes by taking \R{10}$_3$ into account
\bey\nonumber
\sum_{jk}\Big\{(B_{jk}+A_{jk})f^I_k +A_{jk}f^{II}_k\Big\}|\Phi^j><\Phi^j|\ =\
\hspace{2.5cm}
\\ \nonumber
=\ \sum_{jk}\Big\{(B_{jk}+A_{jk})<\Phi^k|k_B \ln(Z\varrho)\Phi^k>+
\hspace{3.5cm}
\\  \label{C10}
+A_{jk}<\Phi^k |\frac{{\cal H}}{\Theta}\Phi^k >\Big\}|\Phi^j><\Phi^j|.
\vspace{.3cm}\eey
The modified von Neumann equation \R{5}$_1$ contains the time derivatives of the
weights of the density operator which are connected with the entropy exchange
\R{9}$_1$ and with the entropy production \R{10}. In conventional quantum mechanics,
the constituitive properties of the system are given by the Hamiltonian which determines
the power exchange \R{5f} with the system's environment. In the semi-classical
description of quantum thermodynamics, the power exchange is supplemented by the
entropy exchange and by the entropy production as an internal quantity of the system,
whereas the Hamiltonian remains undecomposed\footnote{that means: no interaction
term is in the Hamiltonian}. As the power exchange, also the
entropy exchange and the entropy production are constitutive quantities which
are tranfered to the time derivatives of the  weights of the density operator
according to \R{C1} and \R{C2}. Thus, the Hamiltonian and the constitutive mappings
$\bf A$ and $\bf B$ determine the constitutive quantities of semi-classical quantum
thermodynamics: power and heat exchange,
entropy production and contact temperature\footnote{as we will see in the next
section \R{52}} and consequently the entropy exchange.

\section{Temperature as a Quantum Quantity}

Starting with \R{bb41f}$_3$, we obtain by taking \R{9}, \R{2.3a} and \R{41e}$_1$
into consideration
\bee{I5}
\mbox{Tr}\Big(\frac{\cal H}{\Theta}\ro_{ex}\Big)\ =\
-k_B\mbox{Tr}\Big(\ro_{ex}\ln(Z\varrho)\Big).
\ee
This results in a quantum mechanical expression 
for the reciprocal contact temperature of an undecomposed non-isolated closed system
\byy{52}
\frac{1}{\Theta} &=& -\ \frac{\mbox{Tr}\Big(k_B\ro_{ex}\ln(Z\varrho)\Big)}
{\mbox{Tr}({\cal H}\ro_{ex})},
\\ \label{I5a}
\ro_{ex} &=& \sum_{jk}\Big\{A_{jk}<\Phi^k|\Big(k_B \ln(Z\varrho)+
\frac{{\cal H}}{\Theta}\Big)\Phi^k >\Big\}|\Phi^j><\Phi^j|.
\eey
Consequently, axiom IV does not only determine the entropy production
\R{bb41g} and the entropy exchange \R{M1}, but allows to replace
the classical contact temperature by the quantum mechanical expression \R{52}.
\vspace{.3cm}\newline
According to \R{45A}$_{1,2}$, \R{C6} and \R{C7}, we obtain in equilibrium
\bee{I6}
\ro{_{ex}^{eq}}\ =\ 0.
\ee
Consequently, the contact temperature cannot be represented by \R{52} in equilibrium
of non-isolated closed systems for which the equilibrium condition \R{48a} is valid
according to \R{9b} and \R{8a}. The contact temperature is according to \R{8a} not
defined for isolated sytems, because $\st{\td}{Q}$ is identically zero independent of the
environment's temperature $T^\Box$.

\section{Summary}

The von Neumann equation \R{5} is modified by introducing time dependent
weights \R{a5} of the statistical operator generating an additional term, the
propagator \R{5}$_2$ which is traceless and decomposes into two parts \R{41d}, 
an exchange part \R{41e}$_1$ and an irreversibility part \R{41e}$_2$.
This decomposition is caused by the presupposition that the propagator is
sensitive to an isolation of the considered system \R{x41}.
The exchange part determines the entropy exchange \R{M1} between the
undecomposed system and its environment, whereas the irreversibility part
determines the non-negative entropy production \R{bb41g} which is 
insensitive to any isolation of the system. 
\vspace{.3cm}\newline
Starting with the Shannon entropy \R{4}, the time rate of the entropy \R{2.3a} and
also the entropy exchange \R{9} are in contrast to the conventional quantum mechanics
different from zero. Having derived the time rate of the entropy and  the entropy
exchange, the entropy production \R{10} is determined.
\vspace{.3cm}\newline
Because in undecomposed systems the Hamiltonian does not contain an interaction
term, the exchanges between the closed system and its environment are treated
semi-classically by using the power exchange and the entropy exchange which contains  
the system's contact temperature \R{9b}. Because the entropy production is 
insensitive to an isolation of the system, the contact temperature can be represented
quantum mechanically \R{52} by the exchange propagator, the 
Hamiltonian and by the density operator of the non-isolated system. 
\vspace{.3cm}\newline
The equilibrium density operators, \R{a48} and \R{5.1bx}, for closed 
non-isolated and isolated systems
are derived by a thermodynamical induced procedure using a quantum mechanical 
background, but without any concept and use of statistical 
thermodynamics (therefore the headline expression: phenomenological).
Especially, the basic postulate of statistical physics,
the microca\-no\-ni\-cal density operator in isolated systems \R{5.1bx}, is not
postulated here, but is derived by using the modified von Neumann equation
\R{5} and the four axioms I to IV. 
\vspace{.3cm}\newline
Evident is that the semi-classical treatment of quantum thermodynamics represents only
an approximation which needs an extension.
Decomposed Schottky systems, that means system which are described by a Hamiltonian
containing an interaction term, are treated in a second paper: Concepts of Phenomenological
Irreversible Quantum Thermodynamics II: Closed Bipartite Schottky Systems.

\end{document}